The 14th International Conference on Ambient Systems, Networks and Technologies (ANT)
March 15-17, 2023, Leuven, Belgium

# Towards a Deep Learning Pain-Level Detection Deployment at UAE for Patient-Centric-Pain Management and Diagnosis Support: Framework and Performance Evaluation

Leila Ismail[a,b,c,]*, Member, IEEE and Muhammad Danish Waseem[b,c]

[a]*Clouds and Distributed Computing and Systems (CLOUDS) Lab, School of Computing and Information Systems, Faculty of Engineering and Information Technology, The University of Melbourne, Australia*
[b]*Intelligent Distributed Computing and Systems (INDUCE) Research Laboratory, Department of Computer Science and Software Engineering, College of Information Technology, United Arab Emirates University, UAE*
[c]*National Water and Energy Center, United Arab Emirates University, UAE*

**Abstract**

The outbreak of the COVID-19 pandemic revealed the criticality of timely intervention in a situation exacerbated by a shortage in medical staff and equipment. Pain-level screening is the initial step toward identifying the severity of patient conditions. Automatic recognition of state and feelings help in identifying patient symptoms to take immediate adequate action and providing a patient-centric medical plan tailored to a patient's state. In this paper, we propose a framework for pain-level detection for deployment in the United Arab Emirates and assess its performance using the most used approaches in the literature. Our results show that a deployment of a pain-level deep learning detection framework is promising in identifying the pain level accurately.

*Keywords:* Computer Vision; Deep Learning; eHealth; Image Processing; Machine Learning; Pain Detection; Patient-Centric; Smart healthcare

* Corresponding author. Tel.: +971-3-7673333; fax: +971-3-7134343.
*E-mail address:* leila@uaeu.ac.ae

## 1. Introduction

The recent Covid-19 outbreak amplified the need for medical staff to take real-time actions due to the severity of COVID-19 infection and its rapid spread to the respiratory organs, which in many cases can be fatal [1]. Body language is an important way of non-verbal communication to assist in decoding and understanding people's emotions and states. It can involve many parts of the body, such as facial expressions, hands movements, voice tones, and body postures and gestures. In this paper, we focus on automated facial expressions detection that reveals pain intensity levels, as an objective tool to report a patient's pain. Facial expressions coded with Facial Action Units (FACS), such as eyes tightly shut or wrinkles around the nose can provide an objective measure of pain intensity. Automatic pain-level detection can help assess the severity of patients' conditions and take timely actions. It can also aid health allied professionals to put in place a patient-centric approach to pain relief. To date, pain-level detection relies on verbal communication from a patient who may not be able to express the severity of the pain or who speaks a different language than the medical health professional which may add to the ambiguity of the situation.

The United Arab Emirates is one of the top countries leading the deployment of smartness in every aspect of its residents' life. In particular, there is a focus on smart healthcare and smart hospitals for a patient-centric approach and timely intervention [2] [3] [4] [5] [6] [7] [8]. With the emergence of deep learning, several works evaluated the performance of different deep learning algorithms for pain-level detection. However, these algorithms use disparate experimental setups [4-16], making it difficult to achieve an objective comparison among them. To the best of our knowledge, there exists no comprehensive framework that depicts the process of pain-level detection. In this paper, we propose an intelligent deep learning-based pain-level detection framework that would support allied health professionals (doctors, pathologists, therapists, and medical technologists) for better diagnosis and prognosis [9]. This is an initial step towards deploying an end-to-end patient-centric smart hospital system that starts with a smart pain-level screening. The proposed framework is evaluated using the two most used deep learning approaches for pain-level detection in the literature (Table 2), VGG-Face and Resnet, to provide a baseline comparison in a unified setup.

The rest of the paper is organized as follows: In section II, we present a summary of related work. Section III explains our proposed pain detection framework. The experimental setup, experiments, and performance evaluation are discussed in Section IV. Section V concludes the paper.

## 2. Related Work

Table 1. Datasets of Pain Detection from Facial Images and Videos

| Ref | Data Name | Year Published | Dataset Size | Dataset Features |
|---|---|---|---|---|
| [10] | UNBC-McMaster Shoulder Pain | 2011 | 25 participants<br>200 videos, 48,399 total frames | Pain Score (16 levels), FACS coded, AAM Landmarks, VAS, SEN |
| [11] | BioVid Heat Pain | 2013 | 87 participants<br>17,300 5sec videos with 25fps | 4 Pain intensities<br>GSR, ECG, and EMG at trapezius muscle |
| [12] | BP4D-Spontaneous | 2014 | 41 Participants<br>320 2D+3D sequences<br>365,000 total frames 2.6 TB of video | Happiness/Amusement, Sadness, Startle, Embarrassment, Fear, Physical pain, Anger, Disgust<br>FACS coded, 3D and 2D videos |
| [13] | BP4D+ | 2016 | 140 participants<br>10TB+<br>1.4 million frames | Happiness/amusement, Sadness, Startle, Embarrassment, Fear, Physical pain, Anger, Disgust.<br>3D, 2D, thermal, heart rate, blood pressure, skin conductance (EDA), respiration rate, facial features, and FACS codes. |
| [14] | MIntPAIN | 2018 | 20 Participants<br>9366 variable-length videos<br>1,87,939 frames | 5 pain levels<br>RGB + thermal +Depth |

FACs=Facial Action Coding, AAM=Active Appearance Model, VAS=Visual Analogue Scale, SEN=Sensory Scale, GSR = Galvanic Skin Response, ECG= electrocardiogram, EMG= Electromyography, RGB=Red-Green-Blue

Table 2. Work on Deep Learning-based Pain-Level Detection in the Literature.

| Work | Dataset used | Models evaluated | Evaluation metrics |
| --- | --- | --- | --- |
| [15] | UNBC-McMaster Shoulder Pain | Fine-tuned VGG-Face for feature extraction. PCA for dimension reduction. CNN + biLSTM for classification. | Accuracy, AUC, TP, F-score, Precision |
| [16] | UNBC-McMaster Shoulder Pain, BioVid | Used 3D-Resnet model for the spatiotemporal representation of faces in videos. | AUC, MSE |
| [17] | MIntPAIN, UNBC-McMaster Shoulder Pain | Fine-tuned VGG-Face for feature extraction, PCA for feature selection, and an ensemble of shallow 3 CNN + RNN networks for predicting Pain intensity class. | MSE, MAE, Accuracy, AUC, F-score |
| [18] | BioVid | Finetuned VGGFace to distinguish between pain, happiness, and disgust | Precision, Recall, F1 Score |
| [19] | UNBC-McMaster Shoulder Pain | MobileNet, GoogleNet, ResNeXt-50, ResNet18 DenseNet-161 | MSE |
| [20] | UNBC-McMaster Shoulder Pain | Joint training of two custom CNN: One for frame sequence and the second for facial landmarks | MSE, MAE, Accuracy |
| [21] | UNBC-McMaster Shoulder Pain | Used VGG-Face for estimating frame level PSPI and used a fully connected neural network for sequence level pain score with predicted PSPI. | AUC, MAE, Intraclass correlation coefficient |
| [22] | BioVid Heat Pain Dataset | Used Linear Regression, Support Vector Regression, Neural Networks Extreme, and Gradient Boosting. | MAE, RMSE |
| [23] | UNBC-McMaster shoulder pain | Used Attention network for pain intensity estimation. | MSE, Accuracy |
| [24] | UNBC-McMaster shoulder Pain, MIntPAIN | Used VGG-Face and PCA for feature extraction and Temporal Convolution Network for pain intensity classification | AUC, Accuracy, MSE, MAE |
| [25] | AffectNet, UNBC-McMaster shoulder pain | Trained VGG16 on different emotion categories and then did transfer learning for pain recognition. | F1, Precision, Recall, Accuracy |
| [26] | UNBC-McMaster shoulder pain | ResNet-34 and SVM | Acc, Recall, Precision, F1 |
| [27] | UNBC-McMaster shoulder pain, BioVid Heat Pain Database | Used Deit-Transformers for knowledge distillation, GoogleNet | Accuracy |
| **This paper** | **UNBC-McMaster Shoulder Pain** | **Fine-tuned VGG-Face and Resnet-18 to estimate frame-level PSPI** | **Accuracy, MAE, MSE** |

AUC=Area Under Curve, MAE=Mean Absolute Error, RMSE=Root Mean Squared Error, MSE=Mean Squared Error

Table 1 lists the characteristics of publicly available pain detection datasets. Table 2 compares related work on deep learning-based pain-level detection. It shows that these works use different datasets and performance metrics for evaluation. In this paper, we propose an intelligent pain-level prediction framework and evaluate it using the two most used deep learning algorithms in the literature, providing an objective comparison among them.

VGG-Face is widely used as a feature extractor for facial images. [15] extracted the features of the preprocessed images through VGG-face before passing them to the hybrid CNN- BiLSTM classifier. [28] proposed a 2-phased ensemble in which, fine-tuned VGG-face is used in the early phase for feature extraction. [18] fine-tuned VGG-Face for the three-class problem of distinguishing pain from happiness and disgust and applied explainable AI methods to visualize feature explanation. [21] trained VGG-Face to jointly predict frame-level PSPI (Prkachin and Solomon Pain Intensity) and individual facial action units. [25] used HSV color spaces images instead of RGB to extract reduced features from VGG-Face. Several variants of Resnet are also used in the literature. [16] used the 3D variant of Resnet to model spatiotemporal features from videos. [19] compared the performance of several deep convolutional neural networks including, MobileNet, GoogleNet, ResNeXt-50, ResNet18, and DenseNet-16.

## 3. Proposed Deep-Learning Pain-level-Detection Framework

Fig. 1 presents the stages of our proposed automatic pain detection framework that is based on the data analytics lifecycle.

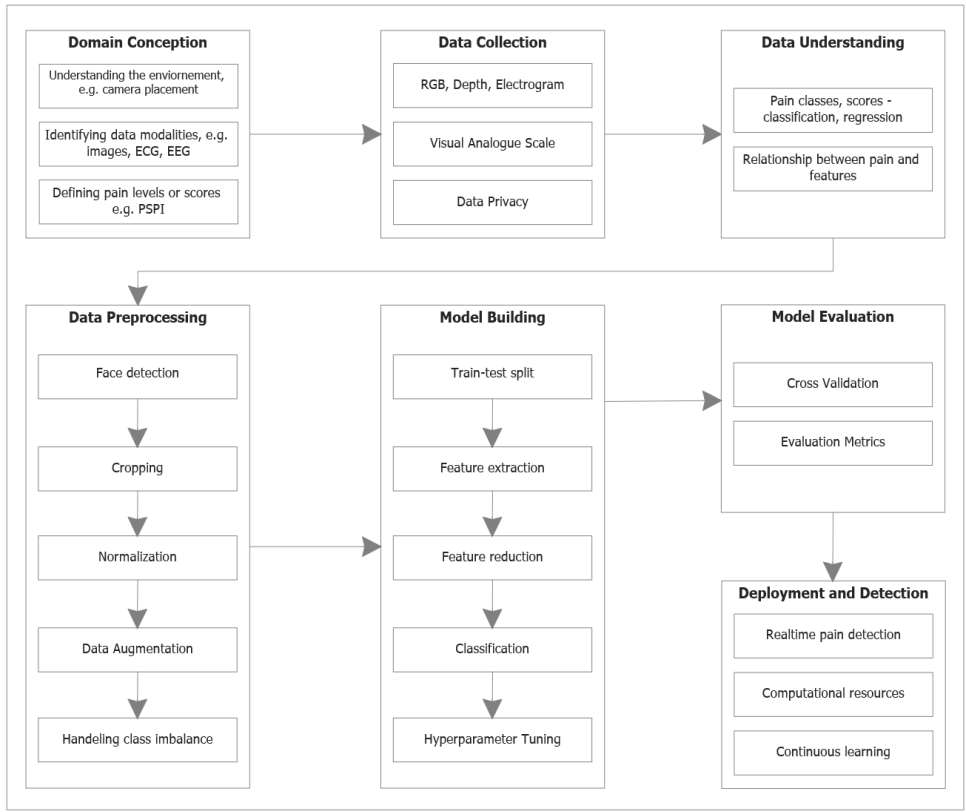

Fig. 1. Stages of the proposed automatic pain detection framework

### 3.1. Domain Conception

In this stage, the problem of automatic pain detection is studied. Pain detection in images can be performed on full body images by analyzing postures or on facial images by examining expressions. For a patient in bed, we can use other features such as ECG along with the camera feed [11]. For accurate pain level detection, it is crucial to understand the ambient environment where the camera is placed. This is because most pain recognition datasets (Table 1) provide frontal face images that are taken in a controlled environment. Close-up facial images may provide more pain-related features than images taken in an uncontrolled environment. However, we must consider the surroundings and varying environment of the inference time which may differ from the training images.

The pain score should be defined as a class label, such as, how many pain levels or range of pain intensities would be modeled. Instead of predicting the pain score directly from an image, we can also predict the Facial Action Units (FACs), which can help us identify other emotions as well in addition to pain. FACs is the coding system for muscle movements that define facial expressions in terms of Action Units (AUs). Table 3 shows examples of AUs. Some AUs have binary values while others have a range of values. The Prkachin and Solomon Pain Intensity (PSPI) [29] is a metric for measuring pain intensity level by simply adding AUs (Eq. 1).

$$Pain\ Score = AU4 + (AU6\ or\ AU7) + (AU9\ or\ AU10) + AU43 \tag{1}$$

Table 3. Example of Facial Action Units

| Action Unit | Descriptor |
| --- | --- |
| AU4 | Brow Lowerer |
| AU6 | Cheek Raiser |
| AU7 | Lid Tightener |
| AU9 | Nose Wrinkler |
| AU10 | Upper Lip Raise |
| AU43 | Eyes Closed |

*3.2. Data Collection and Privacy*

We can collect data ourselves over time or use datasets from public repositories (Table 1). The data collection process depends upon the domain conception step, which defines what features we will be using to train our model. In addition to the RGB data, cameras with depth sensors can provide a 3-dimensional understanding of the faces and environment [14]. Pain level can be associated with the frames in 2 ways: 1) As a classification task that classifies pain intensities in several classes such as no pain, less pain, and high pain [11] or 2) as a regression task such as a range of pain level from 0 to 15 [10]. Pain intensity can be assessed by the Visual Analogue Scale (VAS) [Fig. 2] reported by an expert or the patient.

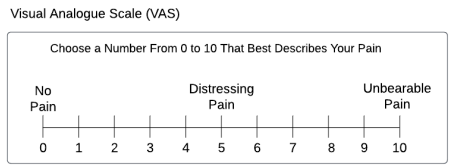

Fig. 2. Visual Analogue Scale for assessing pain intensity.

Patient sensitive data is collected via informed consent and other data privacy-preserving systems can be used [30]. Datasets shown in Table 1 can be used for academic research upon signing an agreement.

*3.3. Data Understanding*

Table 1 shows the publicly available data sets related to pain detection. These datasets provide different features and class labels. For example, [10] provides 16 levels of pain intensities, while [11] has only 4 levels. The pain detection data often suffers from the class imbalance problem since the pain is only induced on high intensities for a very short time, therefore most frames that have a pain score of zero or very low pain. To predict the sequence-level pain score, the video is divided into smaller chunks and the pain score is estimated for the sequence of images. The number of frames per second to process should be selected, based on available computing resources.

*3.4. Data Preprocessing*

In this stage, the dataset should be pre-processed for training. It consists of the following steps:

- *Face detection and cropping*: To avoid background noise, we can crop the faces from the images. Off-the-shelf methods such as DNN or Haar-Cascade face detector from OpenCV [31] can be used to detect faces from the images. After face detection, the face is cropped from the image, which will not only reduce the size of the image but also provide the most relevant features to be modeled.
- *Image Transformations*: Most pre-trained models require an input size of 244 x 244 x 3. Therefore, the images need to be transformed into the size required by the model. Normalization of RGB values can also be performed.
- *Class Imbalance*: When a dataset is imbalanced, over-sampling and under-sampling techniques can be used to reduce the adverse effects of the class imbalance problem [32].
- *Data Augmentation*: Several image transformations such as jitter, center crop, random erasing, etc. can be used for data augmentation, which helps the model generalizes better on unseen data [33].

*3.5. Model Building*

The preprocessed data is used to build the deep learning model. The dataset is divided into training, validation, and testing parts. The validation set is used for hyperparameter tuning and testing set for the evaluation of the model on the unseen data. Optimal hyperparameters' values are used for building the model to improve the prediction performance. At this stage, feature extraction and classification are performed. Several works [15], [18], [21], [24], [28] use VGG-Face, which is a CNN architecture for feature extraction, and pass these features to another neural network or linear machine learning model, such as SVM, for classification. Before passing the features to the classifier, feature selection and reduction techniques such as Principal Component Analysis (PCA) can also be applied [15], [24], [28]. An end-to-end CNN model can also be trained for feature extraction and classification by adding a linear layer at the end of the network for classification. This linear layer will output the vector with dimension as the number of classes. This output is passed to a Softmax layer for obtaining the class probabilities. The class with the highest probability will be considered as the predicted class.

*3.6. Model Evaluation*

To evaluate a created model, the k-fold cross-validation or percentage split approach can be used. Since the number of subjects is generally limited in the pain datasets (Table 1), several works use leave-one-out or k-fold cross-validation [19], [28], [34]. During the training, the loss function can be weighted by the number of observations in each class in the dataset to give more weight to the higher pain intensities. Similarly, during the testing, evaluation metrics can be weighted as well to give more weights to the minority classes i.e., higher pain levels. MAE, MSE, F1-score, Precision, Recall, and Accuracy are widely used as evaluation metrics for pain intensity estimation.

*3.7. Model Deployment*

The developed validated model is used for automatic pain detection by allied healthcare professionals. Resources required for real-time pain detection from videos need to be considered. Continuous learning of the model can be done with supervised or unsupervised learning. However, the model should be evaluated periodically to avoid performance degradation over time.

## 4. Experiments and Performance Evaluation

To evaluate the performance of our framework, we implemented and compared the 2 most used deep learning algorithms, i.e., VGG-Face and Resnet.

Table 4. Layers in VGG-16, Resnet-18, and Resnet-34

| Layer | VGG-16 | Resnet-18 | Resnet-34 |
|---|---|---|---|
| Conv | | 7 x 7, 64, stride 2 | 7 x 7, 64, stride 2 |
| Conv | $[3 \times 3, 64] \times 2$ | $\begin{bmatrix} 3 \times 3, 64 \\ 3 \times 3, 64 \end{bmatrix} \times 2$ | $\begin{bmatrix} 3 \times 3, 64 \\ 3 \times 3, 64 \end{bmatrix} \times 3$ |
| Conv | $[3 \times 3, 128] \times 2$ | $\begin{bmatrix} 3 \times 3, 128 \\ 3 \times 3, 128 \end{bmatrix} \times 2$ | $\begin{bmatrix} 3 \times 3, 128 \\ 3 \times 3, 128 \end{bmatrix} \times 4$ |
| Conv | $[3 \times 3, 256] \times 3$ | $\begin{bmatrix} 3 \times 3, 256 \\ 3 \times 3, 256 \end{bmatrix} \times 2$ | $\begin{bmatrix} 3 \times 3, 256 \\ 3 \times 3, 256 \end{bmatrix} \times 6$ |
| Conv | $[3 \times 3, 512] \times 3$ | $\begin{bmatrix} 3 \times 3, 512 \\ 3 \times 3, 512 \end{bmatrix} \times 2$ | $\begin{bmatrix} 3 \times 3, 512 \\ 3 \times 3, 512 \end{bmatrix} \times 3$ |
| FC | 4096-d FC | | |
| FC | 4096-d FC | | |
| FC | 1000-d FC | 1000-d FC | 1000-d FC |
| **FLOPS** | $1.55 \times 10^{10}$ | $1.8 \times 10^9$ | $3.6 \times 10^9$ |
| **Parameters** | 134.7 M | 11.4 M | 21.5 M |

Conv = Convolutional Layer, FC = Fully Connected Layer,
FLOPS = Floating Point Operations, M = Million

*VGG-Face* [35]: It has the same architecture as VGG16 [36], having 16 convolutional and fully connected layers. However, the difference between VGG16 and VGG-Face is that VGG-Face has been trained on a large-scale facial image dataset. Since VGG-Face is pre-trained on face images it is widely used as a feature extractor for facial images.

*Resnet* [37]: As neural networks become deeper their performance start to degrade due to vanishing gradient problem. Resnet uses skip connections between one or more layers which resolves the issue of vanishing gradients, which allows us to train deeper models. There are different variants of Resnet based on the number of layers used, such as, Resnet-18 having 18 layers and Resnet-34 having 34 layers. Table 4 shows the architecture of these models along with the number of FLOPS and parameters.

We used UNBC-McMaster shoulder pain [10] dataset in which a camera was placed facing the patients having pain. Table 5 shows the hyperparameters we have used for model training. We have used 2 Nvidia RTX-5000 GPUs with total memory of 32 GB. The models were trained to predict frame level PSPI score.

Table 5. Value(s) of Hyperparameters used in the Literature and our Experiments for the Algorithms Under Study.

| Algorithm | Hyperparameters | Value(s) used in the literature | Value(s) used in our experiments |
|---|---|---|---|
| VGG-Face | Optimizer | Adam [12] | Adam Optimizer |
| | Cross Validation | 10-Fold [14], 5-fold [18] | Learning Rate = 0.001 |
| | Epochs | 50 [14] | 5-Fold Cross Validation |
| Resnet | Optimizer | SGD [13], Adam [16] | 100 epochs |
| | Cross Validation | Leave-one-out [16] | Batch size = 256 |
| | Epochs | 200 [13] [16], 5,20,30 [23] | |
| | Batch Size | 64 [13], 8,32,128 [23] | |
| | Learning rate | 0.01 [13] | |

For data pre-processing, we have used 2 OpenCV [31] Haar-Cascade face detectors for getting bounding boxes for faces from raw frames. One for frontal face images and one for side face images. If the face is not detected by the frontal face detector, then we pass it to side face detector. Images were then cropped to only retain the face part and remove the background. We resized the cropped images into 244 x 244 x 3. We used Random Horizontal Flip for data augmentation. Lastly, before passing the image to the model we normalized the images.

To validate the models under study, 5-fold cross-validation is used. The dataset contains total 25 subjects. In each fold we take 15 subjects for training, 5 for validation, and left 5 for testing. We used validation set for evaluating the model during training. If the MAE does not improve for 20 epochs, we did early stopping to avoid overfitting. The model from epoch having minimum MAE was used for testing. We evaluated the performance of the models in terms of MAE, MSE, and Accuracy. We performed end-to-end training of both the models with weighted Cross-Entropy Loss. Weighted loss is used to mitigate the effects of class imbalance on the models. The weights are calculated with Eq. 2.

$$Weights\ of\ Class\ c = \left(\frac{1}{samples\ of\ Class\ c}\right)\left(\frac{Total\ Samples}{Number\ of\ Classes}\right) \quad (2)$$

Table 6 shows the results from VGG-Face, Resnet18 and Resnet34. The average Accuracy, MAE, and MSE for 5 folds is reported.

Table 6. Results of VGGFace and Resnet18 on UNBC-McMaster Dataset

| Model | MAE | MSE | Accuracy |
|---|---|---|---|
| VGG-Face | 0.3589 | 1.7273 | 82.14 |
| Resnet18 | 0.4073 | 2.5002 | 87.89 |
| Resnet34 | 0.5521 | 1.7727 | 78.04 |

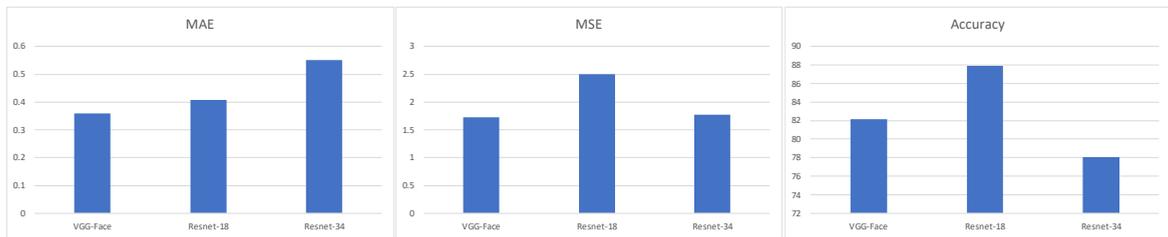

Fig. 3. MAE, MSE, and Accuracy comparison of VGG-Face, Resnet18, and Resnet34

As shown in Fig. 3, Resnet18 has the highest accuracy among the algorithms under study. However, VGG-Face has lowest MAE and MSE, which show how far our predicted pain scores are from the true values. This means on average, the wrongly predicted scores by Resnet18 are further from the true value as compared to the predictions of VGG-Face and Resnet34. MSE penalizes large errors more than MAE. MSE of Resnet18 is higher than VGG-Face and Resnet34, meaning that a patient who is in pain would not be immediately treated, impacting the patient health. Resnet-34 has a greater number of layers, but its accuracy is lower than the other 2 models. It shows that increasing the model size does not necessarily increase the performance. A larger model may tend to overfit given the limited amount of data.

## 5. Conclusion

We presented a framework for pain detection from images and videos to contribute to the UAE vision of smart healthcare and smart hospitals enabled by big data analytics and artificial intelligence. We evaluated the framework on two widely used deep learning models in the literature i.e., VGG-Face and Resnet. Our results show that this framework can provide health practitioners a preliminary checkup metric with very good accuracy. Accuracy should not be the sole performance metric. MSE must be considered, as increase in MSE would mean the patients with high pain levels could be neglected. In future work, we plan to deploy our framework in UAE hospitals emergency units.

**Acknowledgment**
This research was funded by the National Water and Energy Centre of the United Arab Emirates University (Grant 31R215).